\documentclass[12pt,showpacs,showkeys,amsmath,aps,prc]{revtex4}

\usepackage{amsfonts} 
\usepackage{mathrsfs}
\usepackage{subfigure}
\usepackage{supertabular}
\usepackage{color,graphicx}
\usepackage[bookmarksopen]{hyperref}

\def  \a    {\alpha}
\def  \b    {\beta }

\def  \p    {\pi}

\def  \m    {\mu}
\def  \n    {\nu}

\def  \th   {\theta}

\def  \veps {\varepsilon}

\def  \gf   {\gamma^5}

\def  \del  {\partial}

\def  \sls  {\!\!\!/}

\def  \bef  {\begin{figure}}
\def  \eef  {\end{figure}}
\def  \be   {\begin{equation}}
\def  \ee   {\end{equation}}
\def  \ba   {\begin{array}}
\def  \ea   {\end{array}}
\def  \bea  {\begin{eqnarray}}
\def  \eea  {\end{eqnarray}}
\def  \beq  {\begin{eqnarray}}
\def  \eeq  {\end{eqnarray}}
\def  \nn   {\nonumber}
\def  \bd   {\begin{displaymath}}
\def  \ed   {\end{displaymath}}
\def  \bse  {\begin{subequations}}
\def  \ese  {\end{subequations}}
\def  \bwt  {\begin{widetext}}
\def  \ewt  {\end{widetext}}

\def  \ba   {{\bf{a_1}}}

\begin{document}
\title{ Spin dependent Fermi Liquid parameters and properties of polarized quark matter }
\author {Kausik \surname {Pal}}
\email {kausik.pal@saha.ac.in}
\author{Subhrajyoti \surname {Biswas}}
\author {Abhee K. \surname {Dutt-Mazumder}}
\affiliation {High Energy Physics Division, Saha Institute of Nuclear Physics,
 1/AF Bidhannagar, Kolkata 700064, India.}

\medskip

\begin{abstract}
We calculate the spin dependent Fermi liquid parameters (FLPs), 
single particle energies and  energy densities of various
spin states of polarized quark matter. 
The expressions for the incompressibility($K$) 
and sound velocity ($c_1$) 
in terms of the spin dependent FLPs and polarization parameter $(\xi)$ are
derived. Estimated values of $K$ and $c_1$ reveal that the equation of state (EOS) of the 
polarized matter is stiffer than the unpolarized one. Finally we investigate the possibility of the spin polarization (ferromagnetism) phase transition. 
\end{abstract}
\vspace{0.08 cm}

\pacs {04.40.Dg, 12.38.Bx, 12.39.-x, 14.70.Dj, 26.60.-c, 97.60.Jd}

\keywords{Ferromagnetism, Quark matter, Landau parameters.}

\maketitle

\section{Introduction}

One of the important research areas of the contemporary high energy physics has been the
study of matter under extreme conditions. Such a matter, in the laboratory can be produced
by colliding heavy ions at ultra-relativistic energies. Due to asymptotic freedom
of quantum chromodynamics (QCD), it is predicted that the hadronic matter at high temperature and/or density
can undergo a series of phase transitions like confinement-deconfinement and/or chiral phase 
transition {\cite{hwa_book,niegawa05}}. In the high density regime QCD predicts the existence of color superconducting 
state {\cite{rajagopal,inui07,nakano03}}. These apart, the possibility of spin polarized quark liquid {\em i.e.} the 
existence of ferromagnetic phase in dense quark system has also been suggested recently with which 
we are presently concerned {\cite {niegawa05,tatsumi00}}. The properties of dense quark system are particularly relevant
for the study of various astrophysical phenomenon.

The part of the motivation to study the ferromagnetic phase transition in dense quark matter (DQM), 
as mentioned in \cite{tatsumi00} is provided by the discovery of `magnetars' {\cite{tatsumi01}} 
where an extraordinarily high magnetic field
 $\sim 10^{15} G$ exists \cite{tatsumi00,son08}. In \cite{tatsumi00}, it is argued, that the origin of such a high 
magnetic field can be attributed to the existence of spin polarized quark matter 
{\cite{tatsumi06}}. To examine the possibility of ferromagnetism in DQM in 
ref.\cite{tatsumi00} a variational calculation is 
performed where it is observed that there exists a critical density below which spin polarized
quark matter is energetically favorable than unpolarized state. Subsequently various other
calculations were also performed to investigate this issue 
{\cite{son08,niegawa05,inui07,nakano03,tatsumi01,tatsumi06,ohnishi07}}. 
For example, in \cite{nakano03} it is
shown that there is no contradiction between color superconductivity and ferromagnetism and
both of these phase can co-exist. In \cite{ohnishi07}, the same problem was studied in the large
$N_c$ and $N_f$ limit while keeping $N_c/N_f$ fixed where it was shown that spin polarized
state can exist, however, in presence of magnetic screening, color superconductivity or dense 
chiral waves disappear. It might be mentioned that such screening is now 
supported by the lattice calculation {\cite{hand06, ohnishi07}}. 
In{\cite{nakano03}} it is analytically shown that, if quarks are massless, 
ferromagnetism does not appear which is consistent with the conclusion 
drawn in \cite{ohnishi07}. Ref.\cite{son08} shows that ferromagnetism 
might appear in quark matter with Goldstone boson current 
where the magnetization is shown to be related to triangle anomalies.  

In the present work, we apply relativistic Fermi liquid theory (RFLT) to study the possibility of
para-ferro phase transition in DQM. 
The relativistic Fermi liquid theory was developed by
Baym and Chin \cite{baym76} where it has been shown how the  various physical quantities
like chemical potential ($\mu$), incompressibility ($K$), sound velocity ($c_1$) etc. 
can be expressed 
in terms of the Landau parameters (LPs) calculated relativistically. However, the formalism developed 
in \cite{baym76} is valid for unpolarized matter and LPs calculated there are spin averaged.

In this paper we extend the formalism of RFLT and the required LPs are 
calculated by retaining their explicit spin dependencies.  As a result, here various combination of parameters like $f_{0,1}^{++} $, $f_{0,1}^{+-}$, 
$f_{0,1}^{-+} $ and 
$f_{0,1}^{--}$  corresponding to scattering involving up-up, up-down, down-up or down-down spins are appear \cite{baym76}.  Once determined, these parameters are  used to calculate quantities like chemical potentials for the spin up and spin down quarks or the total energy density of the system as a function of 
${\xi}= (n^+_q-n^-_q)/n_q$ and $n_q$ together with various other quantities as we shall see. Here $n_q^+$ and $n_q^-$ correspond to densities of spin up, down quarks respectively and $n_{q}=n_{q}^{+}+n_{q}^{-}$, denotes total quark density 
\cite{tatsumi00}.
We, also compare some of our results with those presented in \cite{tatsumi00} where more direct approach was adopted to calculate the total energy density from the loop. In addition, the present work is extended further to estimate incompressibility and sound velocity in dense quark system for a given fraction of spin-up or down quarks.

Furthermore, in dealing with the massless gluons, we find that naive
series expansion fails and one has to use hard density loop (HDL) corrected gluon propagator to
get the finite result for the LPs involving scattering of like spins {\cite{tatsumi08}}. 
This however does not cause
any problem for the calculation of various physical quantities like chemical potential, exchange
energy, incompressibility etc. We shall see, even though  $f_0$ and $f_1$ (suppressing spin indices) individually remain 
divergent, what appears in our case is the particular combination of these parameters where such divergences cancel.  

The plan of the paper is as follows. In Sec.II, as mentioned before, we extend the formalism of RFLT to include explicit spin dependence. In Sec.III, we derive spin dependent LPs due to one gluon exchange (OGE) for polarized quark matter. Subsequently, we calculate chemical potential and energy density. We find the density dependence of incompressibility $(K)$ and first sound velocity $(c_1)$ with arbitrary spin polarization $(\xi)$. To compare with ref.{\cite{tatsumi00}}, we present ultra-relativistic and non-relativistic results and studied para-ferro phase transition of quark matter. Sec.IV is devoted to summary and conclusion. In Appendix, we calculate various LPs for unlike spin states of scatterer.

\vskip 0.4in
\section {Formalism}

In FLT total energy density $E$ of an interacting system is the functional of 
occupation number $n_{p}$ of the quasi-particle states of momentum $p$. 
The excitation of the system is equivalent to the change of occupation 
number by an amount $\delta n_{p}$. The corresponding energy density of the system is given by {\cite{baym_book,baym76}},
\beq{\label {total_energy}}
E&=&E^{0}+\sum_{s}\int\frac{d^3{p}}{(2\pi)^3}
\varepsilon_{ps}^{0}\delta n_{ps}
+\frac{1}{2}\sum_{ss'}\int\frac{d^3{p}}{(2\pi)^3}\frac{d^3{p'}}{(2\pi)^3}
f_{ps,p's'}
\delta n_{ps}\delta n_{p's'},
\eeq 
where $E^0$ is the ground state energy density and $s$ is the spin index, 
and the quasi-particle energy can be 
written as,
\beq\label{quasi_energy}
\veps_{ps}=\veps_{ps}^{0}+\sum_{s'}\int\frac{d^3{p'}}{(2\pi)^3}f_{ps,p's'}
\delta n_{p's'},
\eeq
where $\veps_{ps}^{0}$ is the non-interacting single particle energy.
The interaction between quasi-particles is given by $f_{ps,p's'}$, which is 
defined 
to be the second derivative of the energy functional with respect to 
occupation functions,
\beq\label{quasi_interac}
f_{ps,p's'}=\frac{\delta^{2}E}{\delta{n}_{ps}~\delta{n}_{p's'}}. 
\eeq

Since, the quasiparticles are well defined only near the Fermi surface,
one assumes
\beq
\veps_{ps}&=&\mu^{s}+v_{f}^{s}(p-p_{f}^{s}).
\eeq

In FLT, the interaction parameter, $f_{ps,p's'}$, is expanded on the basis 
of Legendre polynomials, $P_l$ \cite{baym_book,baym76}. The coefficients of this expansion are known as FLPs, which are given by 

\beq\label{landau_para}
f_l^{ss'}=(2l+1)\int\frac{d\Omega}{4\pi}P_{l}(\cos\theta)f_{ps,p's'},
\eeq

where $\theta$ is the angle between $p$ and $p'$, both taken to be on the Fermi 
surface, and the integration is over all directions of $p$ {\cite{baym76}}. 
Note that unlike \cite{baym_book,baym76}, here we retain explicit spin indices without performing spin summation. We restrict ourselves for $l\le 1$ i.e. $f_{0}^{s}$ and $f_{1}^{s}$, since higher $l$ contribution decreases rapidly as the scattering is dominated by the small angles and the series converges, here, 
$f_{l}^{s}=\frac{1}{2}\sum_{s'}f_{l}^{ss'}$ {\cite{mig_book}}.

The Landau Fermi liquid interaction $f_{ps,p's'}$ is related to the
two particle forward scattering amplitude via {\cite{baym_book,baym76}},

\beq\label{int_para}
f_{ps,p's'}&=&\frac{m_{q}}{\veps_{p}^0}\frac{m_{q}}{\veps_{p'}^0}
{\cal M}_{ps,p's'},
\eeq

where $m_{q}$ is the mass of the quark and the Lorentz invariant matrix  
${\cal M}_{ps,p's'}$ consists of the usual direct and exchange amplitude, 
which may, therefore be evaluated by conventional Feynman rules.
The dimensionless LPs are defined as $F_{l}^{s}=N^{s}(0)f_{l}^{s}$ {\cite{baym76}}, where $N^s(0)$ is the density of states at the Fermi surface is given by,

\beq\label{dens_of_state} 
N^{s}(0)&=&\int\frac{\rm d^3{p}}{(2\pi)^3}
\delta(\veps_{ps}-\mu^{s})\nn\\
&=&\frac{g_{deg}p_{f}^{s^2}}{2\pi^2}\left(\frac{\del p}{\del\veps_{ps}}
\right)_{p=p_{f}^{s}}\nn\\
&\simeq&\frac{g_{deg}p_{f}^{s}\veps_{f}^{s}}{2\pi^2}.
\eeq
 Here $g_{deg}$ is the degeneracy factor. In our case $g_{deg}=N_{c}N_{f}$ where
$N_{c}$ and $N_{f}$ are the color and flavor index for quark matter. For spin up $(+)$ and spin down $(-)$ quark, density of states will be change accordingly.
In the above expression $(\del p/\del\veps_{ps})_{p=p_{f}^{s}}$ is the inverse Fermi velocity $(1/v_{f}^{s})$ related to the FL parameter $F_{1}^{s}$,

\beq\label{inv_vel}
\frac{1}{v_{f}^{s}}=(\del p/\del\veps_{ps})_{p=p_{f}^{s}}=(\mu^{s}/p_{f}^{s})
(1+F_{1}^{s}/3).
\eeq

With Eq.(\ref{dens_of_state}) and Eq.(\ref{inv_vel}) one reads the general relation as {\cite{matsui81}} 

\beq\label{relative}
\veps_{f}^{s}=\mu^{s}(1+\frac{1}{3}F_{1}^{s}).
\eeq
 
The compression modulus or incompressibility $(K)$ of the system is defined by the second derivative of 
total energy density $E$ with respect to the number density $n_{q}$, is given by 
\cite{matsui81,chin77,brown02,negele_book,aguirre07}

\beq\label{compres1}
K&=&9n_{q}\frac{\del^2 E}{\del n_{q}^2}.
\eeq

Now we introduce a polarization parameter ${\xi}$ by the equations, 
$n_{q}^{+}=n_{q}(1+{\xi})/2$ and 
$n_{q}^{-}=n_{q}(1-{\xi})/2$  under the condition $0\le{\xi}\le 1$ 
{\cite{tatsumi00}}. The Fermi momenta in the spin-polarized quark matter then are $p_{f}^{+}=p_{f}(1+{\xi})^{1/3}$ and 
$p_{f}^{-}=p_{f}(1-{\xi})^{1/3}$, where $p_{f}=(\pi^2n_{q})^{1/3}$, is the 
Fermi momentum of the unpolarized matter $({\xi}=0)$. So, there are two Fermi surfaces corresponding to spin-up $(+)$ and spin-down $(-)$ states, such that 
$E\equiv E(n_{q}^{+},n_{q}^{-})$. We have

\beq\label{delE_deln}
\frac{\del E}{\del n_{q}}&=&\frac{\del E}{\del n_{q}^{+}}
\frac{\del n_{q}^{+}}{\del n_{q}}+\frac{\del E}{\del n_{q}^{-}}
\frac{\del n_{q}^{-}}{\del n_{q}}\nn\\
&=&\frac{1}{2}\left[(1+{\xi})\mu^{+}+(1-{\xi})\mu^{-}\right]
\eeq
Using Eq.(\ref{delE_deln}), the incompressibility becomes{\cite{aguirre07}}
\beq\label{compres}
K&=&\frac{9n_{q}}{4}\left[(1+{\xi})^2\frac{\del \mu^{+}}{\del n_{q}^{+}}
+(1-{\xi})^2\frac{\del \mu^{-}}{\del n_{q}^{-}}\right]\nn\\
&=&\frac{9n_{q}}{4}\left[(1+{\xi})^2\left(\frac{1+F_{0}^{+}}{N^{+}(0)}\right)
+(1-{\xi})^2\left(\frac{1+F_{0}^{-}}{N^{-}(0)}\right)\right],
\eeq
where {\cite{baym76}}

\beq
\frac{\del \mu^{s}}{\del n_{q}^{s}}&=&\frac{1+F_{0}^{s}}{N^{s}(0)}.
\eeq

Similarly, the relativistic first sound velocity is given by the first derivative of pressure $P$ with respect to energy density $E$. Since 
$P={\sum}_{s}\mu^{s}n_{q}^{s}-E$ {\cite{aguirre07, matsui81}}, we have,
 
\beq\label{soundvel}
c_{1}^{2}=\frac{\del P}{\del E}
&=&\frac{\del P}{\del n_{q}}\frac{\del n_{q}}{\del E}\nn\\
&=&\left[\frac{(1+{\xi})n_{q}^{+}\frac{\del \mu^{+}}{\del n_{q}^{+}}
+(1-{\xi})n_{q}^{-}\frac{\del \mu^{-}}{\del n_{q}^{-}}}
{(1+{\xi})\mu^{+}+(1-{\xi})\mu^{-}}\right]\nn\\
&=&\frac{n_{q}}{2[(1+{\xi})\mu^{+}+(1-{\xi})\mu^{-}]}
\left[(1+{\xi})^2\left(\frac{1+F_{0}^{+}}{N^{+}(0)}\right)
+(1-{\xi})^2\left(\frac{1+F_{0}^{-}}{N^{-}(0)}\right)\right].
\eeq
In the above Eq.(\ref{compres}) and Eq.(\ref{soundvel}), 
$N^{\pm}(0)$ and $F_{0}^{\pm}$ corresponds to density of states at Fermi surface and dimensionless LP for spin up $(+)$ and spin down $(-)$ quark respectively. For unpolarized matter, $\xi=0$ implying $\mu^+=\mu^-$, $F_0^+=F_0^-$ and 
$N^+(0)=N^-(0)$. From Eq.(\ref{compres}) and (\ref{soundvel}) we have the well known result as $K=9n_{q}\frac{\del \mu}{\del n_{q}}$ {\cite{matsui81}} and 
$c_1^2=\frac{n_{q}}{\mu}\frac{\del\mu}{\del n_{q}}$ {\cite{baym76}}.

\vskip 0.4in
\section{Landau parameters for polarized quark matter}

In this section we calculate LPs for quark matter with explicit spin dependencies. 
We choose spin $s$ along $z$ axis {\em i.e.} $s\equiv (0,0,\pm 1)$ and represent spin-up and down states by their signs. 
For a four-dimensional description of the
polarization state, it is convenient to define a 4-vector $a^{\mu}$
which, in the rest frame of each quark, is same as the three-dimensional 
vector $s$; since $s$ is an axial vector, $a^{\mu}$ is a
4-pseudovector. This 4-vector is orthogonal to the 4-momentum in the
rest frame (in which $a^{\mu}=(0, s), P^{\mu}=(m_{q},0)$); in
any frame we therefore have $a^{\mu}P_{\mu}=0$ 
{\cite{tatsumi00,tatsumi05,lifs_book}}.

The components of the 4-vector $a^{\mu}$ in a frame in which the
particle is moving with momentum $p$ are found by a Lorentz
transformation from the rest frame {\cite{lifs_book}},
\beq
a &=& s+\frac{p(s\cdot p)}{m_{q}(\veps_{p}+m_{q})};~~a^{0}=\frac{p\cdot s}{m_{q}}
\eeq
 with $\veps_{p}=\sqrt{p^2+m_{q}^2}$. We can define projection
 operator $P(a)$ on each of spin polarization,
 $P(a)=\frac{1}{2}(1+\gf a\sls )$. Accordingly the polarization density
 matrix $\rho$ is given by the expression
\beq\label{pola_dens_mat}
\rho(P,s)&=&\frac{1}{2m_{q}}(P\sls+m_{q})P(a),
\eeq
which is normalized by the condition, ${\rm Tr}\rho(P,s)=1$. 
The mean value of the spin is then given by the quantity {\cite{lifs_book}}
\beq
 s_{av}&=&\frac{1}{2}\frac{m_{q}}{\veps_{p}}{\rm
  Tr}(\rho\gamma_{0}\Sigma)
=\frac{1}{2}\frac{m_{q}}{\veps_{p}}{\rm Tr}
(\rho\gamma_{5}{\gamma})\nn\\
&=&\frac{1}{2}\frac{m_{q}}{\veps_{p}}\left(s
+\frac{p(s\cdot p)}{m_{q}(\veps_{p}+m_{q})}\right),
\eeq
which is reduced to $s_{av}=\frac{1}{2}s$ in the
non-relativistic limit.

We consider the color-symmetric forward scattering amplitude of
the two quarks around the Fermi surface by the OGE interaction. The direct term does not contribute as it involves trace of single color matrices like 
${\rm Tr}\lambda_{a}$, which vanishes. Thus the leading contribution comes from the exchange (Fock) term {\cite{tatsumi00}}:

\beq\label{amplitude}
{\cal
  M}_{ps,p's'}^{ex}&=&-\frac{1}{3}\sum_{i}\frac{1}{3}\sum_{j}\left[{\bar
  U_{\beta}(P')g(t^{a})_{ji}\gamma^{\mu}U_{\a}(P)}\right]
\left(\frac{-g_{\m\n}}{(P-P')^2}\right)\left[{\bar
  U_{\a}(P)g(t^{a})_{ij}\gamma^{\nu}U_{\b}(P')}\right]\nn\\
&=&\frac{4}{9}\frac{1}{(P-P')^2}{\rm Tr}[\gamma_{\m}\rho(P,s)
\gamma^{\mu}\rho(P',s')],
\eeq

where $\a$, $\b$ is the flavor level, $i,j$ is the quark color index, 
$t^{a}(=\lambda_{a}/2)$ is the color matrix and $g$ is the coupling constant. Since gluon is flavor blind, the $u-$channel diagrams 
contribute only when $\a=\b$; {\em i.e.} scattering of quarks
with same flavor{\cite{leader_book}}. This means that the Fermi sphere of each flavor makes an independent contribution. Thus the potential energy receives a factor $N_{f}$. On the other hand, the quarks with different colors can take part in the exchange process, giving rise to a factor $N_{c}^2$. Eventually the potential energy density is proportional to $N_{f}N_{c}^2g^2$. For the kinetic energy density, there arises an overall factor $N_{c}N_{f}$. Thus, the factor 
$N_{c}N_{f}$ factorizes out of the total energy density and the competition between the kinetic and potential energies is not influenced by the number of flavor. The number of flavor neither encourages or discourages ferromagnetism 
{\cite{ohnishi07}}.

Without loss of generality, for the calculation of energy density and
other related quantities,
we consider one-flavor quark matter. 
With the help of polarization density matrices given in
Eq.(\ref{pola_dens_mat}), we have from Eq.(\ref{amplitude}) the interaction amplitude as {\cite{tatsumi00}}
\beq\label{int_ampl}
{\cal M}_{ps,p's'}^{ex}
&=&\frac{2g^2}{9m_{q}^2}\frac{1}{(P-P')^2}
[2m_{q}^2-P.P'-(p\cdot s)(p'\cdot s')
+m_{q}^2(s\cdot s')+\frac{1}{(\veps_{p}+m_{q})(\veps_{p'}+m_{q})}
\nn\\&&\times
\{m_{q}(\veps_{p}+m_{q})(p'\cdot s)(p'\cdot s')
+m_{q}(\veps_{p'}+m_{q})(p\cdot s)(p\cdot s')
+(p\cdot p')(p\cdot s)(p'\cdot s')].\nn\\
\eeq

From Eq.(\ref{int_para}) the quasiparticle interaction parameter is given by
\beq\label{int_para1}
f_{ps,p's'}^{ex}&=&\frac{m_{q}}{\veps_{p}}\frac{m_{q}}{\veps_{p'}}
\mathcal{M}_{ps,p's'}^{ex}
\eeq
Here the spin may be either parallel ($s=s'$) or anti-parallel ($s=-s'$). Thus scattering possibilities are denoted by $(+,+)$, $(+,-)$, $(-,-)$ etc. Motivated 
by {\cite{mig_book}}, in analogy with isospin we define spin dependent 
interaction parameter as 
$f^{+}_{pp'}=\frac{1}{2}(f^{++}_{pp'}+f^{+-}_{pp'})$ and 
$f^{-}_{pp'}=\frac{1}{2}(f^{--}_{pp'}+f^{-+}_{pp'})$. Note that, 
$f^{+-}_{pp'}=f^{-+}_{pp'}$.

For $(+,+)$ scattering the interaction parameter is given by

\beq
f^{++}_{pp'\vert{p=p'=p_{f}^{+}}}&=&-\frac{g^2}{9\veps_{f}^{+2}}
\frac{1}{p_{f}^{+2}(1-\cos\th)}[2m_{q}^2-p_{f}^{+2}(1-\cos\th)-p_{f}^{+2}
\cos\th_{1}\cos\th_{2}\nn\\&&
+\frac{1}{(\veps_{f}^{+}+m_{q})^2}\{m_{q}(\veps_{f}^{+}+m_{q})p_{f}^{+2}
(\cos^2\th_{1}+\cos^2\th_{2})
+p_{f}^{+4}\cos\th\cos\th_{1}\cos\th_{2}\}],\nn\\
\eeq
where $\hat{p}\cdot\hat{s}=\cos\th_{1}$ ;
$\hat{p'}\cdot\hat{s}=\cos\th_{2}$ and Fermi energy 
$\veps_{f}^{+}=(p_{f}^{+2}+m_{q}^2)^{1/2}$. Since spin and momentum have no preferred direction, we have done angular average of the spin dependent parameter {\cite{tatsumi07}}: 

\beq\label{ang_av_pp}
{\overline{f^{++}}}_{pp'\vert{p=p'=p_{f}^{+}}}&=&
\int \frac{\rm d\Omega_{1}}{4\pi}\int \frac{\rm d\Omega_{2}}{4\pi}
f^{++}_{pp'\vert{p=p'=p_{f}^{+}}}\nn\\
&=&-\frac{g^2}{9\veps_{f}^{+2}p_{f}^{+2}
(1-\cos\th)}\left[2m_{q}^2-p_{f}^{+2}(1-\cos\th)+\frac{2m_{q}p_{f}^{+2}}
{3(\veps_{f}^{+}+m_{q})}\right].
\eeq

\footnotetext[1]{denoted hereafter ${\overline{f_{pp'}}}= f_{pp'}$.}
$^1$ With the help of Eq.(\ref{landau_para}) along with the Eq.(\ref{ang_av_pp})  one can find LPs, but it is to be noted that $f_{0,1}^{++}$ or $f_{0,1}^{--}$ are individually divergent because of the term, $(1-\cos\th)$, in the denominator of the interaction parameter. This divergence disappear if one uses Debye screening mass for gluons or equivalently use HDL corrected gluon propagator while evaluating the scattering amplitudes {\cite{tatsumi07,tatsumi08}}. 
Note that the combination 
$\left(f_{0}^{++(--)}-\frac{1}{3}f_{1}^{++(--)}\right)$ is, however, 
finite as in this case the divergences cancel and we do not calculate
the LPs separately. It would, however, be interesting to see how do
the results modify if HDL calculations are performed to evaluate 
$f_{0,1}^{++(--)}$, $f_{0,1}^{+-}$ and the corresponding physical quantities. 
The numerical estimates suggest that for the results what  we present here, 
the effect of HDL corrections are expected to be small.

From Eq.(\ref{landau_para}), 

\beq\label{f0f1pp}
f_{0}^{++}-\frac{1}{3}f_{1}^{++}&=&-\frac{g^2}{18\veps_{f}^{+2}p_{f}^{+2}}
\int_{-1}^{+1}
\left[2m_{q}^2-p_{f}^{+2}(1-\cos\th)+\frac{2m_{q}p_{f}^{+2}}{3(\veps_{f}^{+}+m_{q})}
\right]{\rm d(\cos\th)}\nn\\
&=&-\frac{g^2}{9\veps_{f}^{+2}p_{f}^{+2}}\left[2m_{q}^2-p_{f}^{+2}
+\frac{2m_{q}p_{f}^{+2}}{3(\veps_{f}^{+}+m_{q})}\right].
\eeq

The above combination will appear in the calculation of the chemical potential and other relevant quantities. For $(+,-)$ scattering, the angular averaged interaction parameter yields

\beq\label{ang_av_pm}
f^{+-}{\Big\vert}_{p=p_{f}^{+},p'=p_{f}^{-}}&=&
\frac{g^2}{9\veps_{f}^{+}\veps_{f}^{-}}
\left[1-\left\{\frac{m_{q}p_{f}^{+2}}{3(\veps_{f}^{+}+m_{q})}
+\frac{m_{q}p_{f}^{-2}}{3(\veps_{f}^{-}+m_{q})}\right\}
\times\frac{1}{(m_{q}^{2}-\veps_{f}^{+}\veps_{f}^{-}
+p_{f}^{+}p_{f}^{-}\cos\th)}\right].\nn\\
\eeq

It is to be noted that, individual LPs for scattering of unlike spin states
are finite {\em i.e.} free of divergences, in contrast to the case involving scattering of like spin states (For details see Appendix).

\subsection{Chemical potential}

Now we proceed to calculate chemical potential, which, in principle, will be
different for spin-up and spin-down quarks, denoted by  
$\mu^{s}$ with $s~(or~ s') = +,-$ for matter containing unequal densities of up and down quarks. To determine the chemical potential with arbitrary polarization $\xi$, we take the distribution function with explicit spin index $(s ~ or ~ s')$, so that variation of distribution function gives \cite{baym_book,aguirre07,pal08}
\beq\label{spin_dist}
\delta n_{q}^{s}&=& -N^{s}(0)\left[\sum_{s'}f_{0}^{ss'}\delta n_{q}^{s'}
-\delta \mu^{s}\right],
\eeq
where $N^{s}(0)$ is given by the Eq.(\ref{dens_of_state}). The Eq.(\ref{spin_dist}) yields
\beq
\frac{\del \mu^{s}}{\del n_{q}^{s}}&=&\frac{1}{N^{s}(0)}+\sum_{s'}f_{0}^{ss'}
\frac{\del n_{q}^{s'}}{\del n_{q}^{s}}.
\eeq
Separately for spin-up and spin-down states we have

\beq\label{general_mu}
\left(\begin{array}{c}
\del{\mu^+}\\
\del{\mu^-}
\end{array}\right)
&=&\left(\begin{array}{cc}
\frac{1}{N^+(0)}+f_{0}^{++}  &  f_{0}^{+-}\\
     f_{0}^{-+}              &  \frac{1}{N^-(0)}+f_{0}^{--}
\end{array}\right)
\left(\begin{array}{c}
\del{n_{q}^{+}}\\
\del{n_{q}^{-}}
\end{array}\right),\nn\\
\eeq
where the superscripts $++$ and $+-$ denote scattering of quasiparticle with up-up and up-down spin states. For unpolarized matter the upper and lower component become equal which gives rise to the well known result {\cite{baym76}}
\beq\label{mudmu}
\mu{\rm d}\mu&=&\left[p_{f}+\frac{g_{deg}\mu p_{f}^2}{2\pi^2}
(f_{0}-\frac{1}{3}f_{1})\right]{\rm d}p_{f}.
\eeq
In general the chemical potential (both for spin-up and spin-down) is the combination of like and unlike spin states. 
By adjusting the constant of integration {\cite{baym76}}, 
the chemical potential of spin-up quark turns out to be
\beq\label{mup}
\mu^{+}&=&\veps_{f}^{+}-\frac{g^2}{6\pi^2\veps_{f}^{+}}
\left[\frac{11}{6}m_{q}^{2}\ln\left(\frac{p_{f}^{+}+\veps_{f}^{+}}{m_{q}}\right)
+\frac{2}{3}p_{f}^{+}m_{q}-\frac{p_{f}^{+}\veps_{f}^{+}}{2}\right]\nn\\&&
+\frac{g^2}{72\pi^2\veps_{f}^{+}}\left[-\frac{2m_{q}^3}{p_{f}^{+}}
\ln\left(\frac{p_{f}^{+}+p_{f}^{-}}{p_{f}^{+}-p_{f}^{-}}\right)
+\frac{4m_{q}^2\veps_{f}^{+}}{p_{f}^{+}}
\left\{\ln\left(\frac{p_{f}^{+}+p_{f}^{-}}{p_{f}^{+}-p_{f}^{-}}\right)
+\ln\left(\frac{p_{f}^{+}\veps_{f}^{-}+p_{f}^{-}\veps_{f}^{+}}
{p_{f}^{+}\veps_{f}^{-}-p_{f}^{-}\veps_{f}^{+}}\right)\right\}\right.\nn\\&&\left.
-14m_{q}^2\ln\left(\frac{p_{f}^{-}+\veps_{f}^{-}}{m_{q}}\right)
+2m_{q}p_{f}^{-}-3m_{q}p_{f}^{-}
\ln\left(\frac{p_{f}^{+}+p_{f}^{-}}{p_{f}^{+}-p_{f}^{-}}\right)\right.\nn\\&&\left.
-\frac{m_{q}}{p_{f}^{+}}(2m_{q}^2+3p_{f}^{+2})
\ln\left(\frac{p_{f}^{+}\veps_{f}^{-}+p_{f}^{-}\veps_{f}^{+}}
{p_{f}^{+}\veps_{f}^{-}-p_{f}^{-}\veps_{f}^{+}}\right)
+6m_{q}\veps_{f}^{+}\ln\left(\frac{p_{f}^{-}+\veps_{f}^{-}}{m_{q}}\right)
\right.\nn\\&&\left.
+\frac{m_{q}}{p_{f}^{+}}\{2\veps_{f}^{-}(2m_{q}-\veps_{f}^{+})-p_{f}^{-2}\}
\ln\left(\frac{\veps_{f}^{-}\veps_{f}^{+}-m_{q}^2-p_{f}^{-}p_{f}^{+}}
{\veps_{f}^{-}\veps_{f}^{+}-m_{q}^2+p_{f}^{-}p_{f}^{+}}\right)
+6p_{f}^{-}\veps_{f}^{-}\right].
\eeq

In the above equation the term in the first square bracket arises due to the 
scattering of like spin states $(++)$, while the latter comes from the
scattering of unlike spin states $(+-)$.

Similarly, for spin-down quark, one may determine $\mu^{-}$ by 
replacing $p_{f}^{\pm}$ with $ p_{f}^{\mp}$ and
$\veps_{f}^{\pm}$ with $\veps_{f}^{\mp}$ in Eq.(\ref{mup}).

\begin{figure}[htb]
\begin{center}
\resizebox{10.0cm}{8.0cm}{\includegraphics[]{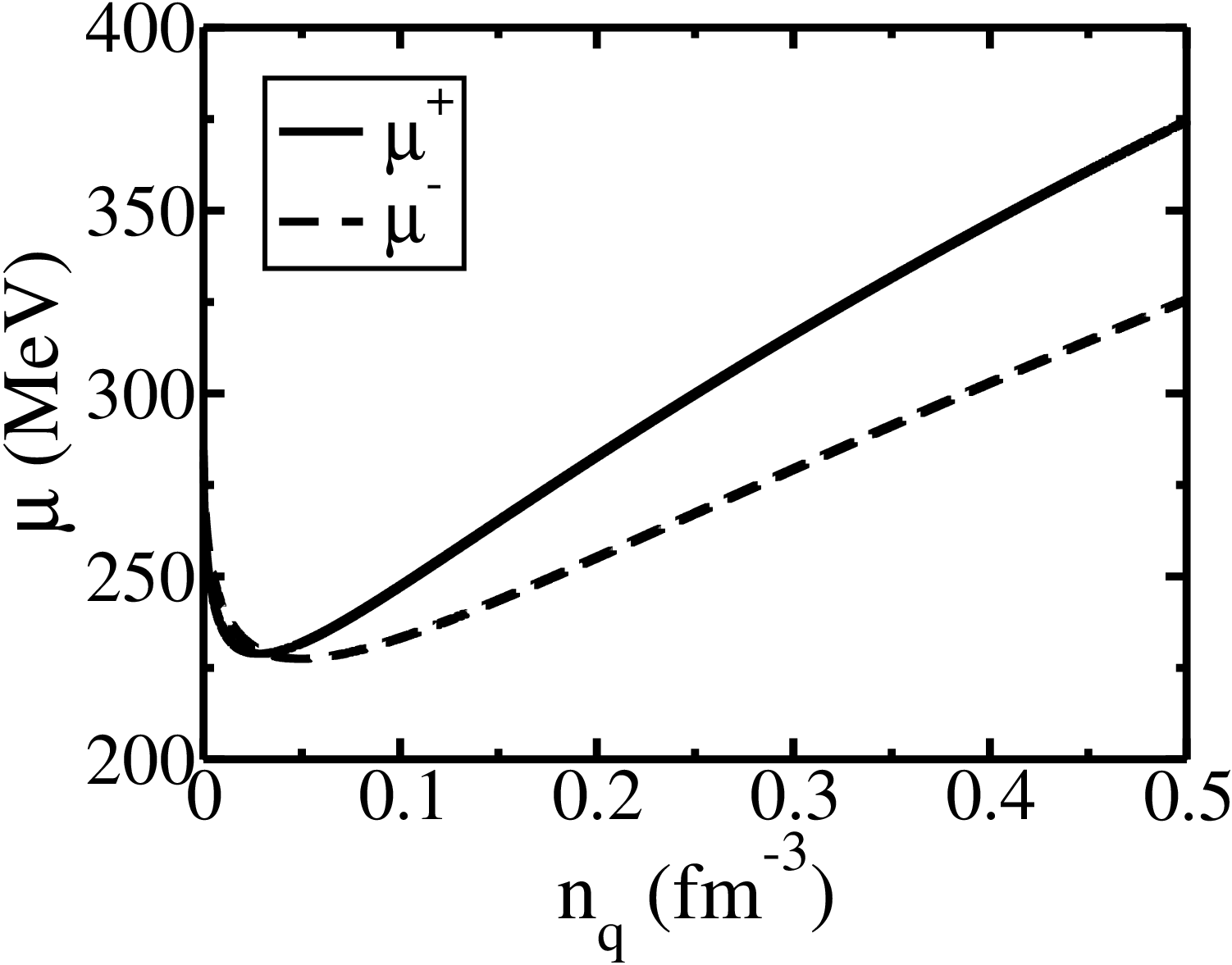}}
\caption{Density dependence of chemical potential of spin-up and spin-down quark denoted by solid and dashed curve respectively.}
\label{mu_p5}
\end{center}
\end{figure}

For the numerical estimation of the above mentioned quantities, following 
ref.{\cite{tatsumi00, degrand75}}, we take $\alpha_{c}=g^2/{4\pi}=2.2$, is the 
fine structure constant of QCD and $m_{q}=300 MeV$. In Fig(\ref{mu_p5}) we 
plot chemical potential for spin-up and spin-down quark as a function of 
density with order parameter $\xi=0.5 $. In real astrophysical calculations, 
the chemical potentials are determined by the $\beta$-equilibrium conditions
where the condition of charge neutrality is also imposed. In Fig(\ref{mu_p5}), 
we, however,
use density $n_q$ and polarization parameter $\xi$ as input parameters 
and  Eq.(\ref{mup}) is used to determine $\mu$ for a system with
one flavor. 

\vskip 0.2in

\subsection{Energy density}

Once the $\mu $ is determined, one can readily calculate the exchange energy density by evaluating \cite{baym76, chin77,pal08}
\beq\label{rel_en_dens}
E_{ex}&=&\int{\rm d}n_{q}(\mu-\veps_{f})
\eeq
After summing up over the color degrees of freedom and evaluating over the Fermi surfaces, we have the exchange energy density. The latter consisting of all types of scattering amplitudes, can be written as  
\beq\label{ex_energy}
E_{ex}&=&E_{ex}^{++}+E_{ex}^{--}+E_{ex}^{+-},
\eeq
which we evaluate numerically. The total kinetic energy density for spin-up 
and spin-down quark is given by
\beq\label{rel_kin}
E_{kin}&=&\frac{3}{16\pi^2}\sum_{s=\pm}
\left[p^{s}_{f}\veps^{s}_{f}(\veps^{s^2}_{f}+p^{s^2}_{f})
-m_{q}^4\ln\left(\frac{\veps^{s}_{f}+p^{s}_{f}}{m_{q}}\right)\right],
\eeq
where $\veps_{f}^{s}=(p_{f}^{s^2}+m_{q}^2)^{1/2}$. The total energy is given by 
the sum of the kinetic energy and the interaction energy $E_{ex}$ {\em i.e.}
\beq\label{eng_total}
E_{tot}&=&E_{kin}+E_{ex}.
\eeq

\begin{figure}[htb]
\begin{center}
\resizebox{9.0cm}{7.0cm}{\includegraphics[]{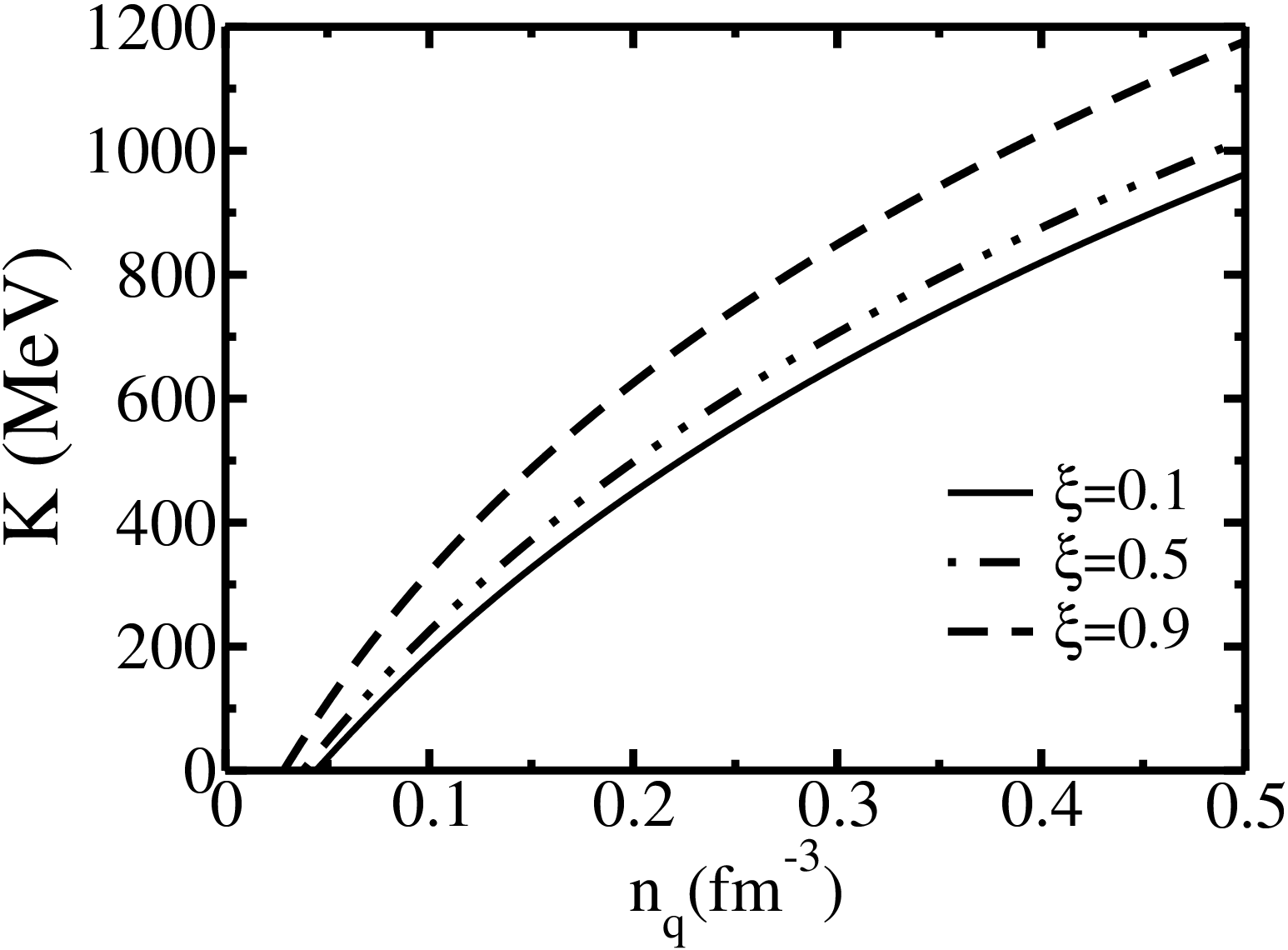}}
\caption{Incompressibility $K$ in quark matter as a function of density for different polarization parameter.}
\label{incompres}
\end{center}
\end{figure}

\begin{figure}[htb]
\begin{center}
\resizebox{9.0cm}{7.0cm}{\includegraphics[]{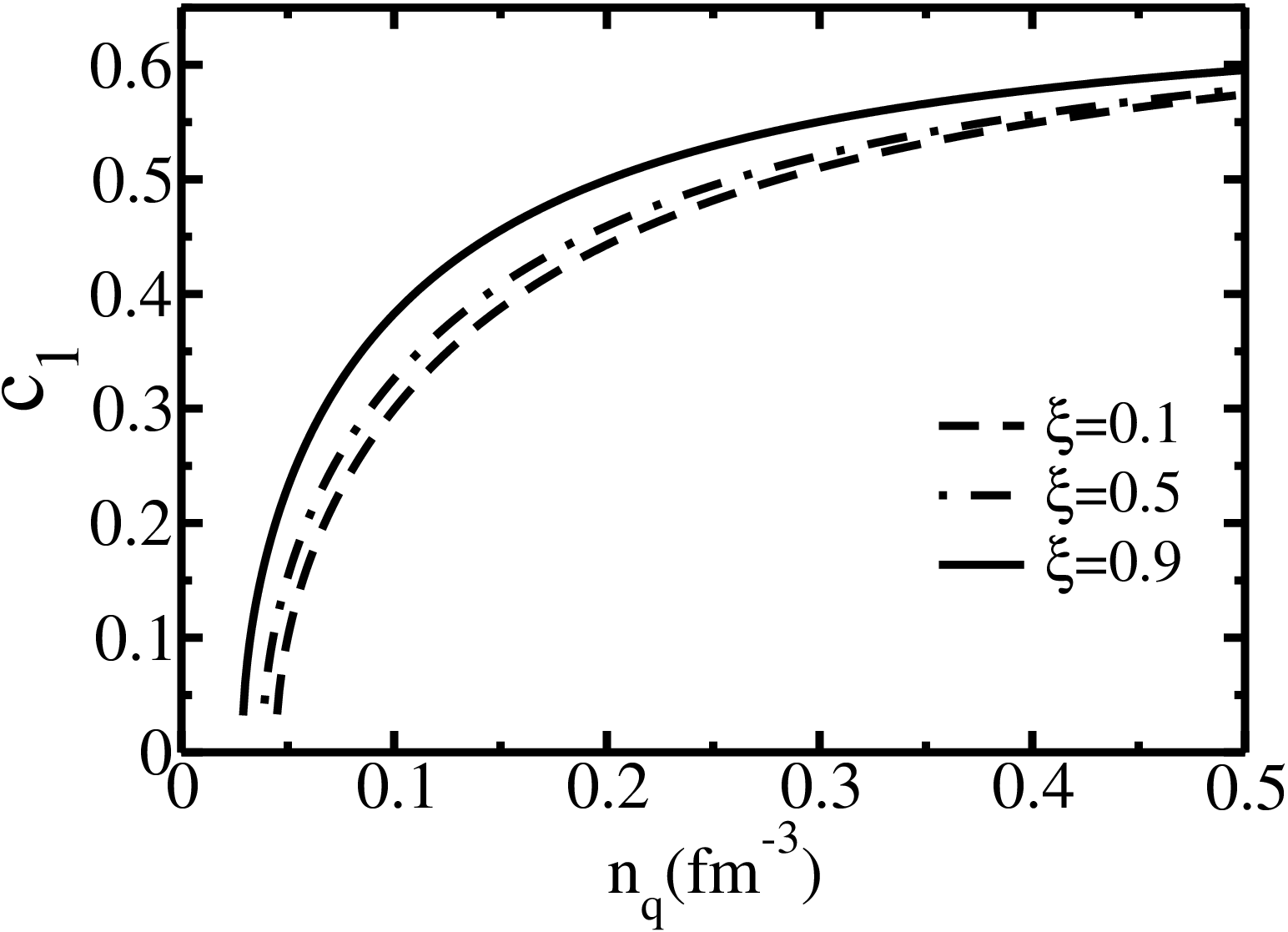}}
\caption{First sound velocity $c_{1}$ in quark matter as a function of density for different polarization parameter.}
\label{first_sound}
\end{center}
\end{figure}

Now we calculate incompressibility and sound velocity by using Eq.(\ref{compres})  and Eq.(\ref{soundvel}). 
In Fig.(\ref{incompres}) and Fig.(\ref{first_sound}) we plot the density dependence of incompressibility and sound velocity. This shows for higher value of the order parameter $\xi$, the incompressibility and the sound velocity becomes higher for the same value of density. Thus the EOS for polarized quark matter is found to be  stiffer than the unpolarized one.

\vskip 0.2in

\subsection{Phase transition}

Bloch first pointed out the possibility of ferromagnetism of electron 
gas where the Fock exchange interaction induces spontaneous spin polarization
{\cite{bloch29}}. Consider the spin polarized electron gas 
interacting by the Coulomb interaction in the background of the positively 
charged ions. Since the direct interaction gives no contribution due to charge 
neutrality, the Fock exchange interaction gives the leading contribution as 
the interaction energy. For spontaneous ferromagnetism, the interaction energy  dominates over the kinetic energy {\cite{tatsumi00,tatsumi00_feb,tatsumi05}}.  

Therefore, if the exchange energy due to OGE interaction 
is negative and becomes greater than the kinetic energy at some density, the quark matter becomes polarized giving rise to ferromagnetism {\cite{tatsumi00}}.

To check whether our results for the total energy density are consistent with 
ref.{\cite{tatsumi00}}, we consider two limiting cases corresponding to the ultra-relativistic (UR) and non-relativistic (NR) regimes.
In the ultra-relativistic (UR) limit, $p_{f}^{s}\gg m_{q}$, then using 
Eq.(\ref{mup}) we have
\beq\label{mupur}
\mu^{+,ur}&=&p_{f}^{+}+\frac{\a_{c}}{3\pi}\left[p_{f}^{+}
+\frac{p_{f}^{-2}}{p_{f}^{+}}\right],
\eeq
Similarly one can find $\mu^{-,ur}$ by replacing $p_{f}^{\pm}$ with $ p_{f}^{\mp}$.

One can arrive at the same expression $\mu^{\pm,ur}$ by taking UR limit of the scattering amplitude. For $(+,+)$ scattering one gets the interaction parameter as 
 
\beq\label{ur_pp}
f_{pp'}^{++,ur}&=&\frac{g^2}{9pp'}
\left(1+\cos\th_{1}\cos\th_{2}\right).
\eeq

After taking angular average of the interaction parameter and with the help of 
Eq.(\ref{landau_para}), we find that $f_{1}^{++}$ vanishes. Thus we have
\beq
f_{pp'}^{++,ur}{\Big|}_{p=p'=p_{f}^{+}}=f_{0}^{++,ur}&=&\frac{g^2}{9p_{f}^{+2}}.
\eeq 
Similarly for $(+,-)$ scattering, the interaction parameter yields
\beq\label{ur_pm}
f_{pp'}^{+-,ur}&=&\frac{g^2}{9pp'}
\left(1-\cos\th_{1}\cos\th_{2}\right).
\eeq
The only existing LP is $f_{0}^{+-}$  and other higher order LPs does not contribute. Hence we get 
\beq
f_{pp'}^{+-,ur}{\Big|}_{p=p_{f}^{+},p'=p_{f}^{-}}=f_{0}^{+-,ur}
&=&\frac{g^2}{9p_{f}^{+}p_{f}^{-}}.
\eeq
It is observed that, in UR limit, all the LPs are finite. Now the chemical 
potential for spin-up quark is found to be
\beq\label{ur_chpot_p}
\mu^{+,ur}&=&p_{f}^{+}+\frac{\a_{c}}{3\pi}\left[p_{f}^{+}
+\frac{p_{f}^{-2}}{p_{f}^{+}}\right].
\eeq
The chemical potential, $\mu^{-,ur}$, can be obtained by replacing 
$p_{f}^{\pm}$ with $p_{f}^{\mp}$ in Eq.(\ref{ur_chpot_p}).

Using Eqs.(\ref{rel_en_dens}) and (\ref{ex_energy}), the exchange energy densities are given by 
\beq\label{eng_ur}
\left.\begin{array}{lll}
E_{ex}^{++,ur}&=\frac{\a_{c}}{8\p^3}p_{f}^{+4}\\
E_{ex}^{--,ur}&=\frac{\a_{c}}{8\p^3}p_{f}^{-4}\\
E_{ex}^{+-,ur}&=\frac{\a_{c}}{4\p^3}p_{f}^{+2}p_{f}^{-2}.
\end{array}
\right\}
\eeq

Thus the final expression for the exchange energy density in the UR limit 
is found to be 
\beq\label{ur_eng_dens}
E_{ex}^{ur}&=&\frac{\a_{c}}{8\pi^3}p_{f}^4\left[(1+{\xi})^{4/3}+
(1-{\xi})^{4/3}+2(1-{\xi}^2)^{2/3}\right].
\eeq
This result is same as in ref.{\cite{tatsumi00}}. 

Similarly from Eq.(\ref{rel_kin}), the kinetic energy density in UR limit
takes the following form {\cite{tatsumi00}}:
\beq
E_{kin}^{ur}&=&\frac{3p_{f}^4}{8\pi^2}
\left[(1+{\xi})^{4/3}+(1-{\xi})^{4/3}\right].
\eeq

In the NR limit, $p (or~p')\ll m_{q}$, the interaction parameter reduces to a simple form, 
\beq\label{nr_int_para}
f^{nr}_{ps,p's'}&=&-\frac{g^2}{9pp'}
\left[\frac{1+s\cdot s'}{(1-\cos\th)}\right].
\eeq
For spin anti-parallel interaction $s=-s'$, then $f_{ps,p's'}^{nr}=0$. Thus the contribution due to the scattering of quarks with unlike spin states vanishes and  the dominant contribution to energy density comes from the parallel spin states $(s=s')$. For the $(s,s)$ scattering, the interaction parameter yields
\beq
f^{nr, s}_{pp'}{\Big|}_{p=p'=p_{f}^{s}}&=&-\frac{2g^2}{9p_{f}^{s^2}(1-\cos\th)},
\eeq
where $s=+$ or $-$ according to scattering process. In NR limit one gets
\beq
(f_{0}^{s}-\frac{1}{3}f_{1}^{s})&=&-\frac{2g^2}{9p_{f}^{s^2}}.
\eeq
The NR chemical potential $\mu^{nr}$ is given by
\beq
\mu^{nr,s}&=&m_{q}-\frac{g^2}{3\pi^2}p_{f}^{s}.
\eeq

Using Eq.(\ref{rel_en_dens}), the exchange energy density for the $(+,+)$ scattering is given by
\beq
E_{ex}^{++,nr}&=&-\frac{g^2}{8\pi^4}p_{f}^{4}(1+\xi)^{4/3}.
\eeq
Similarly for $(-,-)$ scattering, we have
\beq
E_{ex}^{--,nr}&=&-\frac{g^2}{8\pi^4}p_{f}^{4}(1-\xi)^{4/3}.
\eeq
As in the NR limit $E_{ex}^{+-}=E_{ex}^{-+}=0$ as mentioned before, so 
from Eq.(\ref{ex_energy}) the exchange energy density yields
\beq\label{eng_dens_nr}
E_{ex}^{nr}&=&-\frac{\a_{c}}{2\pi^3}p_{f}^4
\left[(1+\xi)^{4/3}+(1-\xi)^{4/3}\right].
\eeq

Thus the energy density, in this limit, becomes negative. The kinetic energy
density turns out to be {\cite{tatsumi00}}
\beq\label{kin_eng_nr}
E_{kin}^{nr}&=&\frac{3p_{f}^5}{20\pi^2m_{q}}
\left[(1+\xi)^{5/3}+(1-\xi)^{5/3}\right].
\eeq
In NR limit ferromagnetism can appear as a consequence of competition between the kinetic energy and the Coulomb potential energy {\cite{ohnishi07}}. The latter favors the spin alignment due to quantum effect. When the energy gain due to the spin alignment dominate over the increase in the kinetic energy at some density, the unpolarized state suddenly turns into the completely polarized state 
{\cite{tatsumi00_apr}}. On the other hand, in UR limit,
the contribution to the energy density not only comes from the like spin states but also unlike spin states of scatterer (see \cite{tatsumi00} for detailed discussion). 

\vskip 0.2in

\begin{figure}[htb]
\begin{center}
\resizebox{9.0cm}{7.0cm}{\includegraphics[]{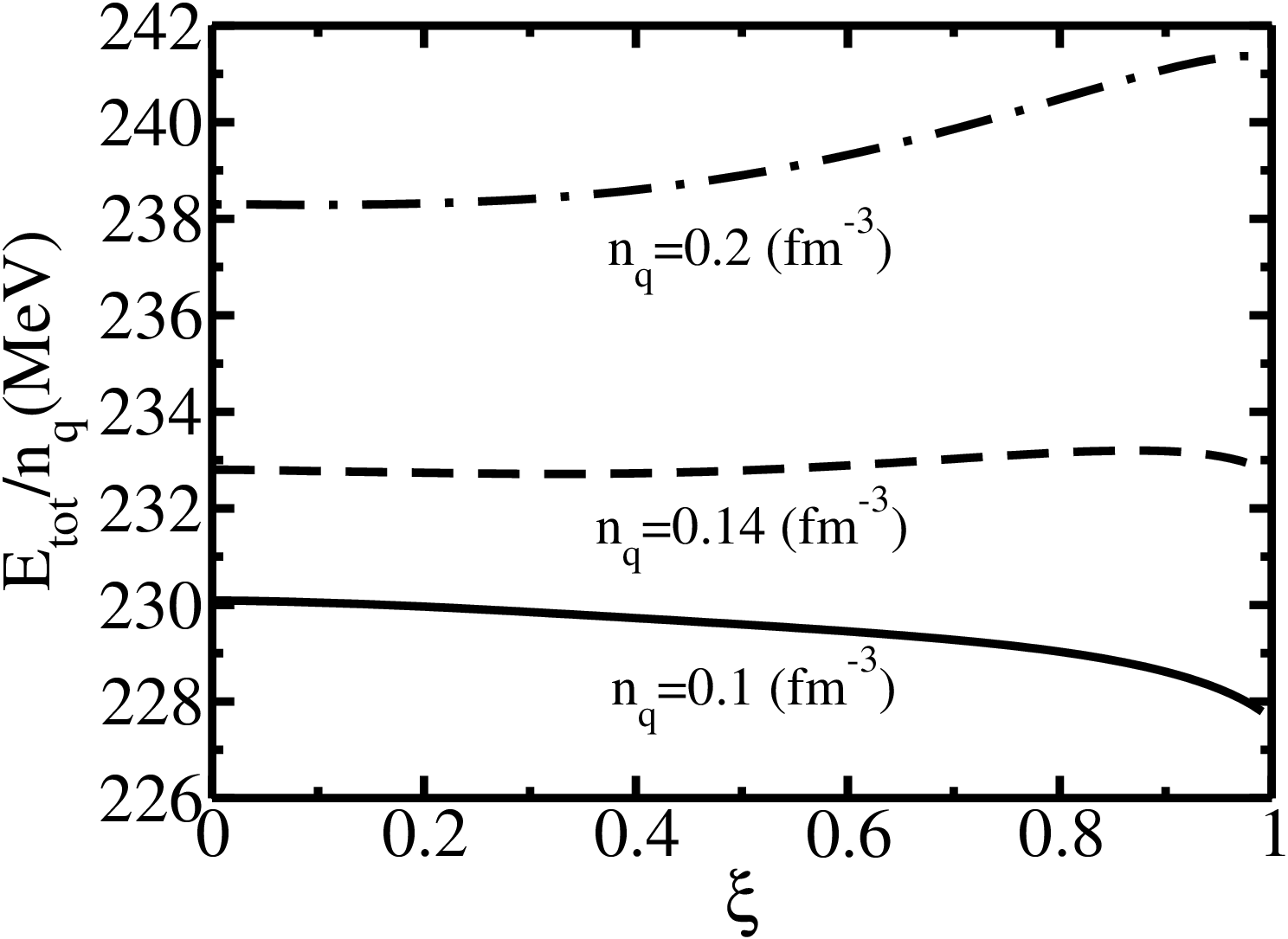}}
\caption{The total energy of quark liquid as a function of polarization parameter at $n_{q}=0.1{~\rm fm^{-3}}$, $n_{q}=0.14{~\rm fm^{-3}}$ and 
$n_{q}=0.2{~\rm fm^{-3}}$. The critical density is found to be 
$n_{q}^c=0.14{~\rm fm^{-3}}$ in this case.}
\label{engtot}
\end{center}
\end{figure}



\begin{figure}[htb]
\centering
\includegraphics[scale=0.3,angle=0]{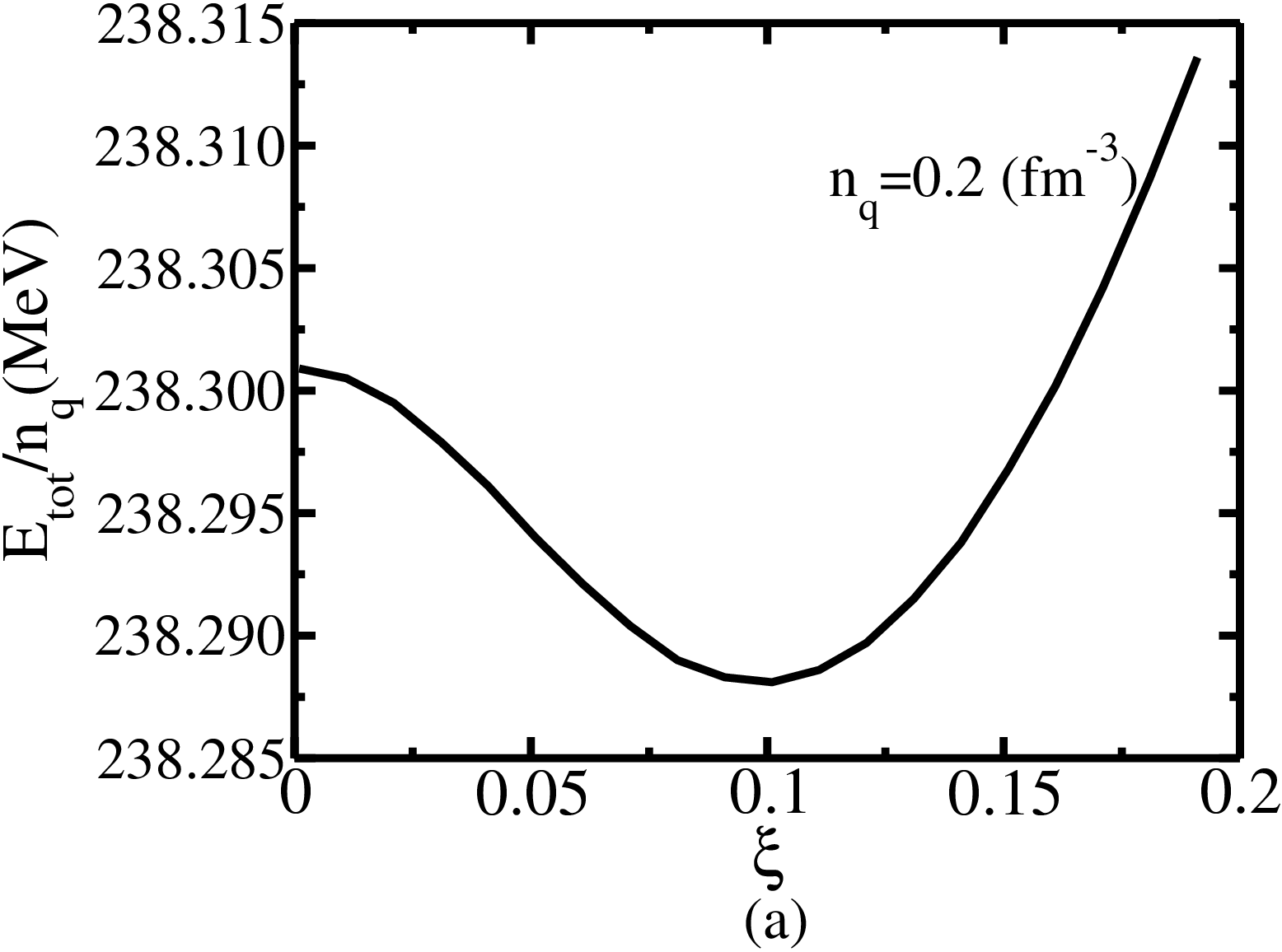}
\hfill
\includegraphics[scale=0.3,angle=0]{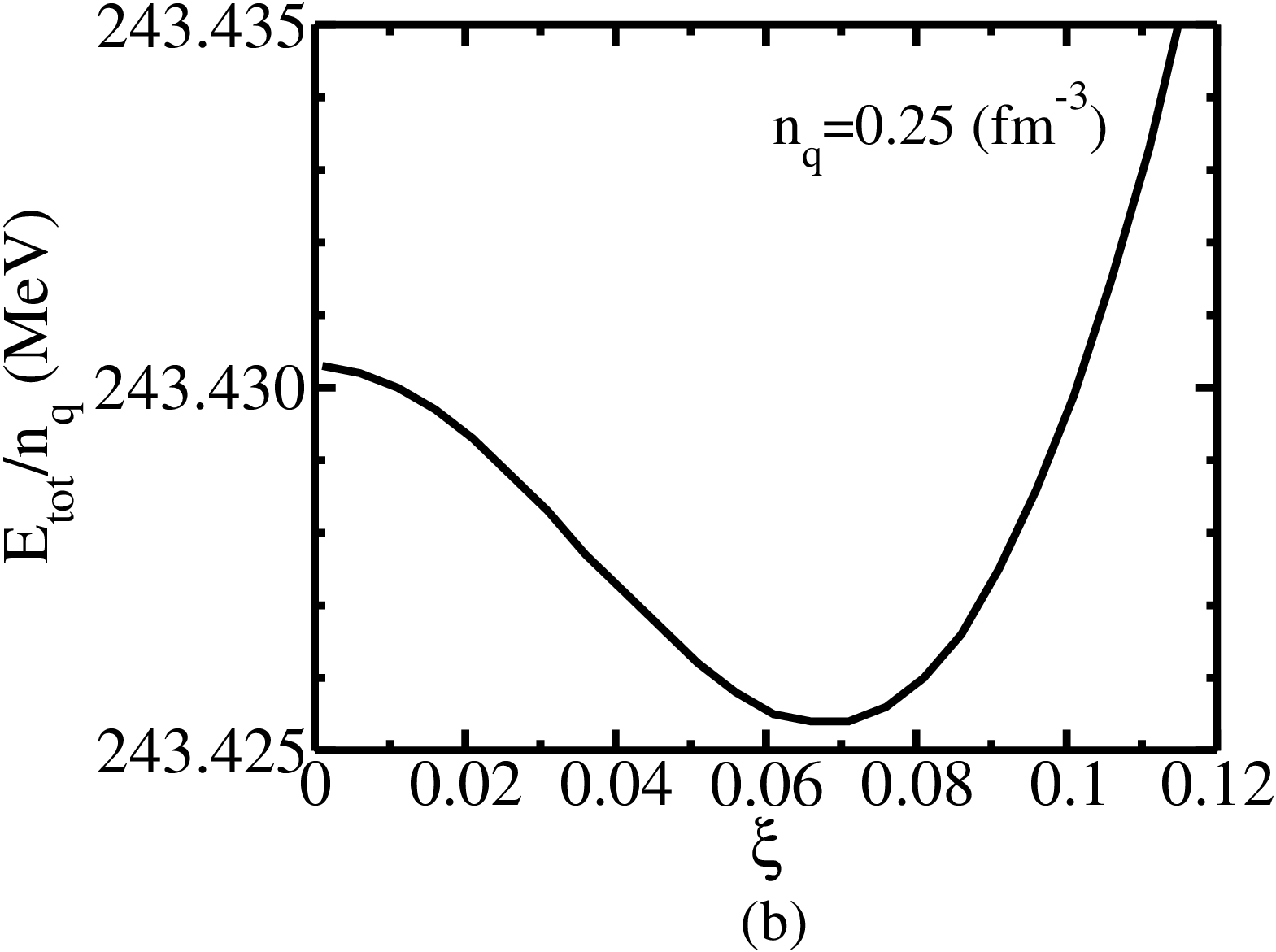}\\

\vskip 0.2in

\includegraphics[scale=0.3,angle=0]{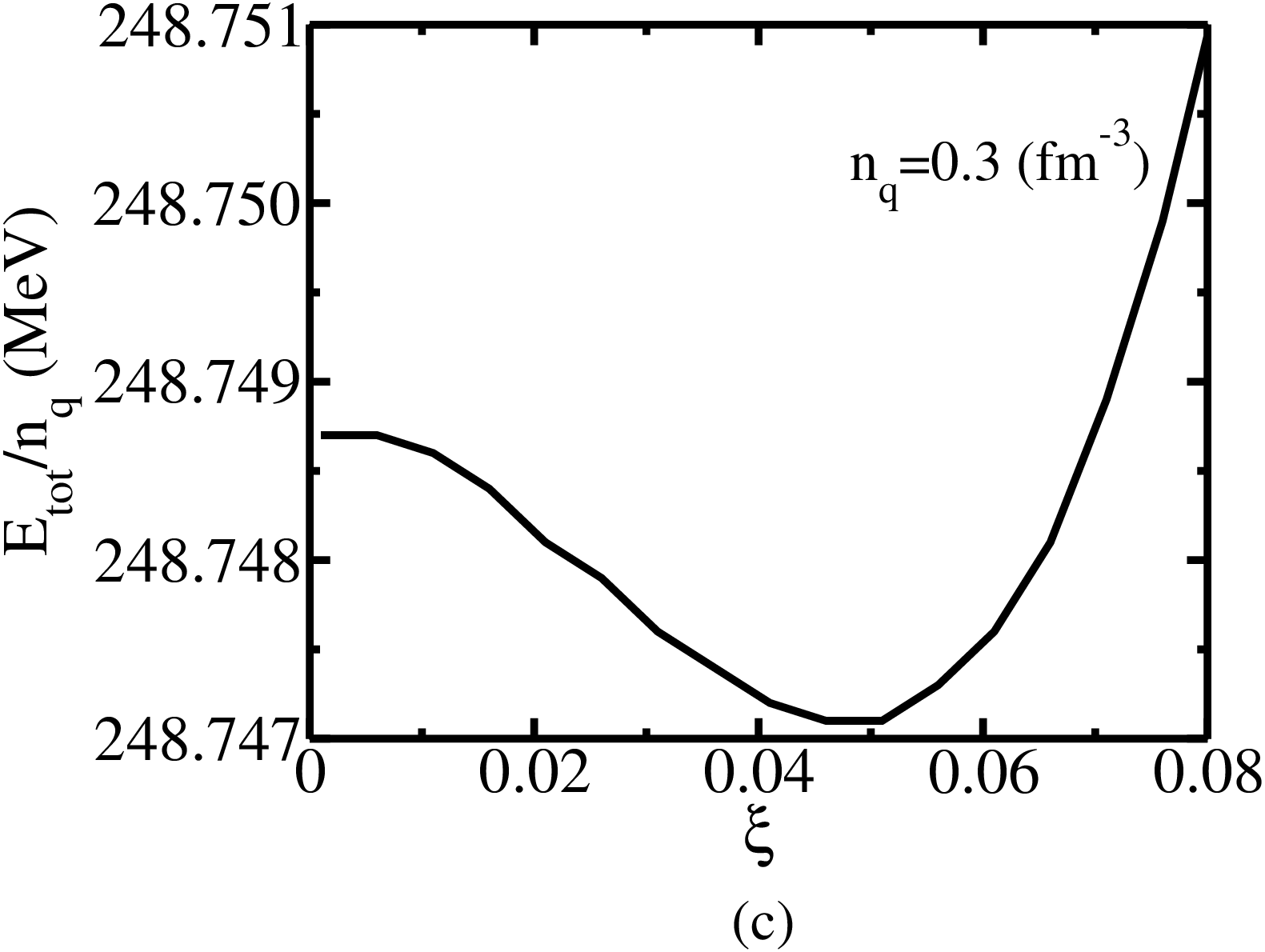}
\includegraphics[scale=0.3,angle=0]{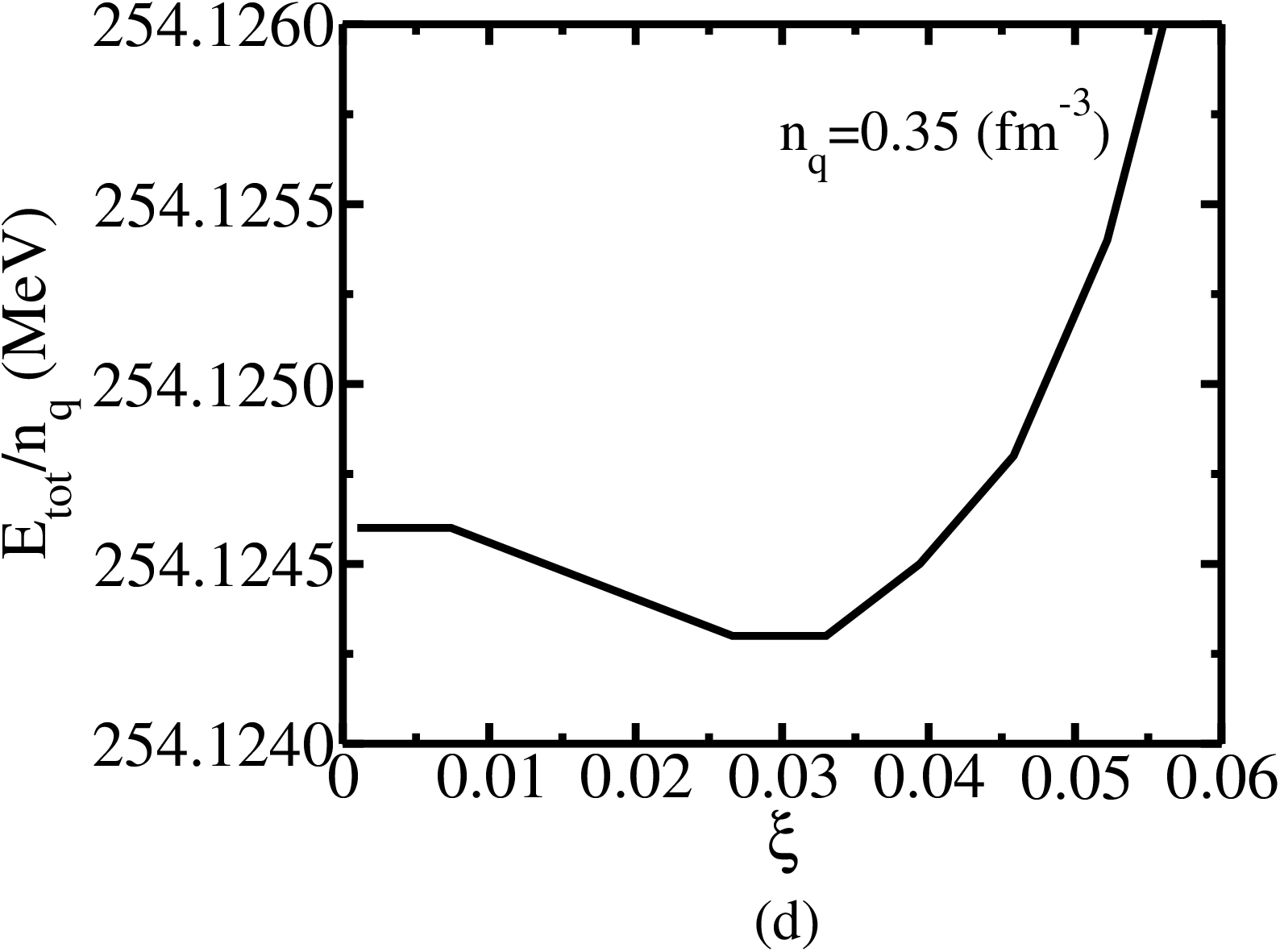}\\
\hfill
\caption{Metastable ferromagnetic state as a function of polarization parameter for
different densities.}
\label{metas}
\end{figure}


To check the consistency we compare our result derived in RFLT approach with 
that of ref{\cite{tatsumi00}} derived from two loop ring diagrams.
In Fig.(\ref{engtot}) we plot $E_{tot}/n_q$ as a function of polarization 
parameter $\xi$. The results clearly show that for lower density 
$(<0.14 {~\rm fm^{-3}})$, total energy favors at $\xi=1$ which indicates  completely polarized state; while at higher density, the system becomes
unpolarized $(\xi=0)$. Thus the polarization parameter suddenly
changes from $\xi=1$ to $\xi=0$ as one increases the number density of the system. So the phase transition is first order and the critical density $n_{q}^{c}$ is around $0.14{~\rm fm^{-3}}$. 

In Fig.(\ref{metas}) we show total energy as a function of polarization parameter for different densities. In every plot, there is a minima which corresponds 
to a possible metastable state. 
We notice that when density increases metastable state arises for lower values of polarization parameter $\xi $. For example, at density $\sim 0.2 {~\rm fm^{-3}}$ minima arises at $\xi=0.1$ while at density 
$\sim 0.35 {~\rm fm^{-3}}$ minima arises at $\xi=0.03$. Thus the metastable state shows a tendency of disappearance as the density increased.

\vskip 0.4in
\section{summary and conclusion}

To summarize and conclude, in this work we have applied RFLT to study the properties of dense 
quark matter. Accordingly, we calculate the FLPs by retaining their explicit spin
dependencies. We also show how the physical quantities like chemical potential of
spin up and spin down states, their energy densities and the quantities like
incompressibility, sound velocity for polarized quark matter can be expressed in terms of these 
spin dependent RFLPs.  For the scattering involving like spin states, the LPs 
$f_{0,1}^{++}$ and $f_{0,1}^{--}$ are found to diverge. However, we show that
 the combination in which they 
appear in the calculation of the physical quantities such divergences cancel. 
For the scattering involving unlike spin states no such divergence appear. The appearance
of such divergences is related to the unscreened gluonic interaction between the quarks
which might be cured by invoking hard dense loop corrected gluon propagator. We do not perform
such calculation here and postpone this for future investigation. 
As far as the equation of state (EOS) is concerned, we in the present model find
that the EOS for the polarized quark matter is stiffer than the unpolarized one.
In addition, we also show that there exists a metastable state which disappear 
at higher density, although it seems that the effect is tiny.

We reconfirm that DQM can exhibit ferromagnetism at low density as was
originally suggested in {\cite{tatsumi00}}. However, the density at which 
the spin polarized ferromagnetic state in the present model might appear
depends strongly on the quark mass. The critical density increases
with increasing mass. In Fig.~(\ref{engtot}), we observe that states 
with $\xi$ appear only below or around normal nuclear density where
deconfined quark matter is not likely to exist. We cannot, however,
ascertain the critical density from the present analysis where we
restrict ourselves only to OGE diagrams and one flavor system. 
In this regime, multi-gluon
exchange processes {\cite{ohnishi07}} might play an important role.
Furthermore, the correlations as given by the ring diagrams can
also change the conclusion. Further work therefore is necessary
to understand the existence of ferromagnetic quark matter in real
multi flavor system which might appear in astrophysics.

\section{Appendix}

In the text the interaction parameter $f_{pp'}^{+-}$ for unlike spin states was calculated. Here we give detail expression of Landau parameters. 
With the help of Eq.(\ref{landau_para}), the LPs are given by,

\beq\label{f0pm}
f_{0}^{+-}&=&\frac{g^2}{18\veps_{f}^{+}\veps_{f}^{-}}
\left[2+\frac{m_{q}[\veps_{f}^{-}p_{f}^{+2}+m_{q}(p_{f}^{+2}+p_{f}^{-2})+
\veps_{f}^{+}p_{f}^{-2}]}{3p_{f}^{+}p_{f}^{-}(m_{q}+\veps_{f}^{+})
(m_{q}+\veps_{f}^{-})}
\ln\left(\frac{m_{q}^2-p_{f}^{+}p_{f}^{-}-\veps_{f}^{+}\veps_{f}^{-}}
{m_{q}^2+p_{f}^{+}p_{f}^{-}-\veps_{f}^{+}\veps_{f}^{-}}\right)\right]\nn\\
\eeq

and 

\beq\label{f1pm}
f_{1}^{+-}&=&\frac{g^2}{18\veps_{f}^{+}\veps_{f}^{-}}
\left[6-\frac{2m_{q}[\veps_{f}^{-}p_{f}^{+2}+m_{q}(p_{f}^{+2}+p_{f}^{-2})+
\veps_{f}^{+}p_{f}^{-2}]}{p_{f}^{+}p_{f}^{-}(m_{q}+\veps_{f}^{+})
(m_{q}+\veps_{f}^{-})}+\right.\nn\\&&\left.
\left(\frac{m_{q}(m_{q}^2-\veps_{f}^{+}\veps_{f}^{-})
[\veps_{f}^{-}p_{f}^{+2}+m_{q}(p_{f}^{+2}+p_{f}^{-2})+\veps_{f}^{+}p_{f}^{-2}]}
{p_{f}^{+2}p_{f}^{-2}(m_{q}+\veps_{f}^{+})(m_{q}+\veps_{f}^{-})}\right)
\ln\left(\frac{m_{q}^2+p_{f}^{+}p_{f}^{-}-\veps_{f}^{+}\veps_{f}^{-}}
{m_{q}^2-p_{f}^{+}p_{f}^{-}-\veps_{f}^{+}\veps_{f}^{-}}\right)
\right].
\nn\\
\eeq

Using Eq.(\ref{f0pm}) and Eq.(\ref{f1pm}) we have
 
\beq\label{f0f1pm}
f_{0}^{+-}-\frac{1}{3}f_{1}^{+-}&=&
\frac{g^2}{18\veps_{f}^{+}\veps_{f}^{-}}
\left[2-\left\{\frac{m_{q}p_{f}^{+2}}{3(\veps_{f}^{+}+m_{q})}
+\frac{m_{q}p_{f}^{-2}}{3(\veps_{f}^{-}+m_{q})}\right\}\times\right.\nn\\&&\left.
\left\{-\frac{2}{p_{f}^{+}p_{f}^{-}}+\frac{(p_{f}^{+}p_{f}^{-}+m_{q}^{2}
-\veps_{f}^{+}\veps_{f}^{-})}{(p_{f}^{+2}p_{f}^{-2})}
\ln\left(\frac{m_{q}^{2}+p_{f}^{+}p_{f}^{-}-\veps_{f}^{+}\veps_{f}^{-}}
{m_{q}^{2}-p_{f}^{+}p_{f}^{-}-\veps_{f}^{+}\veps_{f}^{-}}\right)\right\}
\right].
\eeq

\vskip 0.2in
{\bf Acknowledgments}\\

The authors would like to thank P.Roy for his critical reading of the manuscript.

\end{document}